\documentclass[showpacs,preprintnumbers,amsmath,amssymb]{revtex4}

\usepackage{graphicx}
\usepackage{dcolumn}
\usepackage{bm}

\begin{document}

\preprint{...}

\title{A new form of self-duality equations with topological term }

\author{Yongqiang Wang}
\thanks{Corresponding author}\email{wyq02@st.lzu.edu.cn}
\affiliation{
 Institute of Modern Physics, Chinese Academy of Sciences, Lanzhou
730000, People's Republic of China\\
 Institute of Theoretical
Physics, Lanzhou University,  Lanzhou 730000, People's Republic of
China}
\author{Yuxiao Liu}
\author{Yishi Duan}
\affiliation{ Institute of Theoretical Physics, Lanzhou
University,
 Lanzhou 730000, People's Republic of China}
\date{\today}

\begin{abstract}
   Based on the $U(1)$ gauge potential decomposition theory and
$\phi$-mapping theory, the topological inner structure of the
self-duality (Bogomol'nyi-type) equations are studied. The special
form of the gauge potential decomposition is obtained directly
from the first of the self-duality equations. Using this
decomposition, the topological inner structure of the
Chern-Simons-Higgs (CSH) vortex is discussed. Furthermore, we
obtain a rigorous self-dual equation with topological term for the
first time, in which the topological term has been ignored by
other physicists.

\end{abstract}
 \maketitle

\section{Introduction}
Bogomol'nyi-type vortices and self-dual solutions appear in a
variety of  gauge theories scenarios in $2+1$ dimension, including
Abrikosov-Nielson-Olesen model  \cite{Abrikosov,Nielsen,3},
Chern-Simons-Higgs (CSH) Model \cite{Jackiw, Jackiw2, Hong},
Maxwell-Chern-Simons Higgs Model  \cite{Lee} and Jackiw-Pi model
 \cite{Pi}.

 Let us start with an Abelian
CSH Lagrangian density  \cite{Jackiw}
\begin{equation}\label{000}
L_{CSH}=\frac{1}{4}\kappa\epsilon^{\mu\nu\lambda}A_{\mu}F_{\nu\lambda}+D_{\mu}\phi(D_{\mu}\phi)^{*}-V(\|\phi\|),
\end{equation}
where $\phi$ is the designated charged Higgs complex scalar field
minimally coupled to an Abelian gauge field,
$\frac{1}{4}\kappa\epsilon^{\mu\nu\lambda}A_{\mu}F_{\nu\lambda}$
is the so-called Chern-Simons term, and the covariant derivative
is $D_{\mu}\phi=\partial_{\mu}\phi+ieA_{\mu}\phi$. The first-order
Bogomol'nyi self-duality equations in this model are
\begin{eqnarray}\label{001}
D_{\pm}\phi=0\nonumber \ \ \ \
 (D_{\pm}\equiv D_{1}\pm{i}D_{2}),
 \\B=\pm\frac{2}{\kappa^{2}}\|\phi\|^{2}(\|\phi\|^{2}-\nu^{2}).
\end{eqnarray}

As pointed out by many physicists \cite{Jackiw, 2}, the magnetic
flux of the vortex is
\begin{equation}\label{002}
\Phi=\oint{A_{i}}dx^{i}=\int\epsilon^{ij}\partial_{i}A_{j}dx^{2}=\frac{2\pi{n}}{e},
\end{equation}
here $n$ is a topological index characterizing the vortex
configuration. When the scalar field $\phi$ is decomposed into its
phase and magnitude: $\phi=\rho^{\frac{1}{2}}e^{iw}$, the first of
the self-duality equations (\ref{001}) determines the gauge field:
\begin{equation}\label{103}
A_{i}=-\frac{1}{e}\partial_{i}w\mp\frac{1}{2e}\epsilon_{ij}\partial_{j}\ln\rho.
\end{equation}
The second of Eq. (\ref{001}) reduces to a nonlinear elliptic
equation for the scalar field density $\rho$:
\begin{equation}\label{003}
\bigtriangledown^{2}\ln\rho=\frac{4e}{\kappa^{2}}\rho(\rho-\nu^{2}),
\end{equation}
and this equation is not solvable, or even integrable.

In this paper we accomplish two things. Firstly, using Duan's
$\phi$-mapping theory \cite{4,5,6,7,8}, we study the topological
inner structure of the first Bogomol'nyi self-duality equation and
obtain directly one special form of the gauge potential
decomposition. Using this decomposition, we derive the vortex
configuration given in Eq. (\ref{002}) from the self-duality
equations (\ref{001}), and one sees that the inner structure of
this vortex labelled only by the topological indices of the zero
points of the complex scalar field. Secondly, we study the second
self-duality equation in Eqs. (\ref{001}) and obtain a new scalar
field equation with a topological term, which differs from the
conventional equation (\ref{003}), and the topological term is the
topological current of the vortex. Furthermore, the analytical
solution  of the equation in the special conditions is also
investigated.

\section{U(1) gauge potential decomposition of self-duality equations}
It is well known that the
complex scalar field $\phi$ can be looked upon as a section of a
complex line bundle with base manifold $M$ \cite{Dubrovin}.
Denoting  the charged Higgs complex scalar field $\phi$ as
\begin{equation}\label{004}
\phi=\phi^{1}+i\phi^{2},
\end{equation}
where $\phi^{a}(a=1,2)$ are two components of a two-dimensional
vector field $\vec{\phi}=(\phi^{1},\phi^{2})$ over the base space,
one can introduce the two-dimensional unit vector
\begin{equation}\label{005}
n^{a}=\frac{\phi^{a}}{\|\phi\|},\;\;\;\|\phi\|=(\phi\phi^{\ast})^{\frac{1}{2}}.
\end{equation}

Let us consider the first of self-duality equations (\ref{001})
(firstly, we choose the upper signs):
\begin{equation}\label{006}
D_{+}\phi=0,
\end{equation}
substituting Eq. (\ref{004}) into the above equation, we obtain
two equations
\begin{eqnarray}\label{007}
\partial_{1}\phi^{1}-\partial_{2}\phi^{2}=eA_{1}\phi^{2}+eA_{2}\phi^{1}, \nonumber\\
\partial_{1}\phi^{2}+\partial_{2}\phi^{1}=eA_{2}\phi^{2}-eA_{1}\phi^{1}.
\end{eqnarray}
Making use of the above relations, we derive:
\begin{equation}\label{008}
\partial_{1}\phi^{\ast}\phi-\partial_{1}\phi\phi^{\ast}=2ieA_{1}\|\phi\|^{2}+i(\partial_{2}\phi^{\ast}\phi+\partial_{2}\phi\phi^{\ast}),
\end{equation}
\begin{equation}\label{009}
\partial_{2}\phi^{\ast}\phi-\partial_{2}\phi\phi^{\ast}=2ieA_{2}\|\phi\|^{2}-i(\partial_{1}\phi^{\ast}\phi+\partial_{1}\phi\phi^{\ast}).
\end{equation}
To proceed, we need a fundamental identity-one that appear many
 times throughout our study in the gauge
potential decomposition theory:
\begin{equation}\label{010}
\epsilon_{ab}n^{a}\partial_{i}n^{b}=\frac{1}{2i}\frac{1}{\phi^{\ast}\phi}(\partial_{i}\phi^{\ast}\phi-\partial_{i}\phi\phi^{\ast}),
\end{equation}
using this identity, Eq. (\ref{008}) and Eq. (\ref{009}) become
\begin{equation}\label{011}
eA_{1}=\epsilon_{ab}n^{a}\partial_{1}n^{b}-\frac{1}{2}\partial_{2}\ln(\phi\phi^{\ast}),
\end{equation}
\begin{equation}\label{012}
eA_{2}=\epsilon_{ab}n^{a}\partial_{2}n^{b}+\frac{1}{2}\partial_{1}\ln(\phi\phi^{\ast}).
\end{equation}
Eqs. (\ref{011}) and (\ref{012}) can be rewritten as:
\begin{equation}\label{013}
eA_{i}=\epsilon_{ab}n^{a}\partial_{i}n^{b}-\epsilon_{ij}\frac{1}{2}\partial_{j}\ln(\phi\phi^{\ast}).
\end{equation}
Following the same discussion, we obtain the similar equation from
$D_{-}\phi=0$:
\begin{equation}\label{014}
eA_{i}=\epsilon_{ab}n^{a}\partial_{i}n^{b}+\epsilon_{ij}\frac{1}{2}\partial_{j}\ln(\phi\phi^{\ast}).
\end{equation}
So, from the first self-duality equation $D_{\pm}\phi=0$, we get
\begin{equation}\label{015}
A_{i}=\frac{1}{e}\epsilon_{ab}n^{a}\partial_{i}n^{b}\mp\frac{1}{2e}
\epsilon_{ij}\partial_{j}\ln(\phi\phi^{\ast}).
\end{equation}

The $U(1)$ gauge potential can be decomposed by the Higgs complex
scalar field $\phi$ as
\begin{equation}\label{016}
A_{i}=\beta\epsilon_{ab}\partial_{i}n^{a}n^{b}+\partial_{i}\lambda,
\end{equation}
in which $\beta=\frac{1}{e}$ is a constant and $\lambda=\mp
\frac{1}{2e} \epsilon_{ij}\ln(\phi\phi^{\ast})$ is a phase factor
denoting the $U(1)$ transformation. It is seen that the term
($\partial_{i}\lambda$) in Eq. (\ref{016}) contributes nothing to
the magnetic flux of the vortex (\ref{002}).

From above discussion, it is obvious to see that we obtain one
special form of the general $U(1)$ gauge potential decomposition
from the first self-duality equation,  in the next section, we
will see that the topological inner structure of CSH vortex is
described by gauge potential $A_{i}$ (\ref{015}).

\section{the topological inner structure of CSH vortex}
Based on the decomposition of the gauge potential $A_{i}$
discussed in section II,  we can immediately get the equation of
the magnetic flux of the vortex (\ref{002}). Using the
two-dimensional unit vector field (\ref{005}), we can construct a
topological current:
\begin{equation}\label{017}
J^{\mu}=\epsilon^{\mu\nu\lambda}\partial_{\nu}A_{\lambda}=\frac{1}{e}\epsilon^{\mu\nu\lambda}\epsilon_{ab}\partial_{\nu}n^{a}\partial_{\lambda}n^{b},
\end{equation}
which is the special case of the general $\phi$-mapping
topological current theory \cite{4}. Obviously, the current
(\ref{017}) is conserved. Following the $\phi$-mapping theory, it
can be rigorously proved that
\begin{equation}\label{018}
J^{\mu}=\frac{2\pi}{e}\delta^{2}(\vec{\phi})D^{\mu}(\frac{\phi}{x}),
\end{equation}
where
$D^{\mu}(\frac{\phi}{x})=\frac{1}{2}\epsilon^{\mu\nu\lambda}\epsilon_{ab}\partial_{\nu}\phi^{a}\partial_{\lambda}\phi^{b}$
is the vector Jacobians. This expression provides an important
conclusion: $J^{\mu}=0, \; iff \; \vec{\phi}\neq0; \;
J^{\mu}\neq0, \; iff \; \vec{\phi}=0$. Suppose that the vector
field $\vec{\phi}(\phi^{1},\phi^{2})$ possesses $l$ zeros, denoted
as $z_{i}(i=1,...,l)$. According to the implicit function theorem
\cite{9}, when the zero points $\vec{z_{i}}$ are the regular
points of $\vec{\phi}$, that requires the Jacobians determinant
\begin{equation}\label{021}
D\left(\frac{\phi}{x}\right)\bigg\vert_{z_{i}}\equiv
D^{0}\left(\frac{\phi}{x}\right)\bigg\vert_{z_{i}}\neq{0}.
\end{equation}
The solutions of Eq. (\ref{021}) can be generally obtained:
$\vec{x}=\vec{z_{i}}(t), \; i=1,2, \cdots , l, \; x^{0}=t.$ Using
Eqs. (19) and (\ref{021}), it is easy to prove that
\begin{equation}\label{023}
D^{\mu}\left(\frac{\phi}{x}\right)\bigg\vert_{z_{i}}=D\left(\frac{\phi}{x}\right)\bigg\vert_{z_{i}}\frac{dx^{\mu}}{dt}.
\end{equation}
According to the $\delta$-function theory \cite{10} and  the
$\phi$-mapping theory, one can prove that
\begin{equation}\label{024}
J^{\mu}=\frac{2\pi}{e}\sum_{i=1}^{l}\beta_{i}\eta_{i}\delta^{2}(\vec{x}-\vec{z}_{i})\frac{dx^{\mu}}{dt}\bigg\vert_{z_{i}},
\end{equation}
in which the positive integer $\beta_{i}$ is the Hopf index and
$\eta_{i}= sgn(D(\phi/{x})_{z_{i}})=\pm1$ is the Brouwer degree
\cite{11,5}. Then the density of topological charge can be
expressed as
\begin{equation}\label{025}
J^{0}=\frac{2\pi}{e}\sum_{i=1}^{l}\beta_{i}\eta_{i}\delta^{2}(\vec{x}-\vec{z}_{i}).
\end{equation}
From Eq. (\ref{017}), it is easy to see that
\begin{equation}\label{026}
J^{0}=\epsilon^{ij}\partial_{i}A_{j}.
\end{equation}
So, the total charge of the system given in Eq. (\ref{002}) can be
rewritten as
\begin{equation}\label{027}
Q=\int J^{0}dx^{2}=\frac{2\pi}{e}\sum_{i=1}^{l}\beta_{i}\eta_{i}.
\end{equation}
And the topological index $n$ in Eq. (\ref{002}) has the following
expression
\begin{equation}\label{028}
n=\sum_{i=1}^{l}\beta_{i}\eta_{i}.
\end{equation}
It is obvious to see that there exist $l$ isolated vortices in
which the $i$th vortex possesses charge
$\frac{2\pi}{e}\beta_{i}\eta_{i}$. The vortex corresponds to
$\eta_{i}=+1$, while the antivortex corresponds to $\eta_{i}=-1$.
One can conclude that vortex configuration given in Eq.
(\ref{002}) is a multivortex solution which possesses the inner
structure described by expression (\ref{027}).

\section{self-dual equation with topological term}

The second Bogomol'nyi self-duality equation (\ref{003}) is
meaningless, when the field $\phi=0$. Moreover, no exact solutions
are known for this equation. In this section, based on the
decomposition of $U(1)$, we rewrite the self-dual equation with
topological term and study its general analytical solution in some
condition.

Firstly, based on the special form of the general $U(1)$
decomposition of gauge potential (\ref{015}), we get:
\begin{equation}\label{030}
B=\frac{1}{e}\epsilon^{ij}\epsilon_{ab}\partial_{i}n^{a}\partial_{j}n^{b}\mp\frac{1}{2e}\delta_{jk}\epsilon^{ij}\epsilon^{kl}\partial_{i}\partial_{l}\ln(\phi\phi^{\ast}).
\end{equation}
Substituting above equation into the second Bogomol'nyi
self-duality equation (\ref{001}), we obtain:
\begin{equation}\label{031}
\frac{1}{2}\bigtriangledown^{2}\ln(\phi\phi^{\ast})=\frac{2e}{\kappa^{2}}\|\phi\|^{2}(\|\phi\|^{2}-\nu^{2})\mp\epsilon^{ij}\epsilon_{ab}\partial_{i}n^{a}\partial_{j}n^{b}.
\end{equation}
From Eqs. (\ref{017}) and (\ref{018}), we obtain the
$\delta$-function form of topological term
\begin{equation}\label{041}
\epsilon^{ij}\epsilon_{ab}\partial_{i}n^{a}\partial_{j}n^{b}=eJ^{0}=2\pi\delta^{2}(\vec{\phi})D(\frac{\phi}{x}).
\end{equation}
The second self-dual equation in Eq. (\ref{001}) then can reduce
to a nonlinear elliptic equation for the scalar field density
($\rho=\phi\phi^{\ast}$)
\begin{equation}\label{032}
\bigtriangledown^{2}\ln\rho=\frac{4e}{\kappa^{2}}\rho(\rho-\nu^{2})\mp4\pi\delta^{2}(\vec{\phi})D(\frac{\phi}{x}).
\end{equation}

Comparing with Eq.(\ref{003}), one can find that the conventional
self-dual equation (\ref{003}), in which the topological term has
been ignored, is meaningless when the field $\phi=0$; we get the
self-dual equation with topological term, which is meaning when
the field $\phi=0$. Obviously, topological term is very important
to the inner topological structure of the self-duality equations,
and $\mp4\pi\delta^{2}(\vec{\phi})D(\frac{\phi}{x})$ is the
density of topological charge of the vortex.

  Now the self-dual equation with topological term is more
difficult to be solved, we will study the analytical solution of
the equation in the following condition: 
\begin{equation}\label{033}
\frac{4e}{\kappa^{2}}\rho\nu^{2}\pm4\pi\delta^{2}(\vec{\phi})D(\frac{\phi}{x})=0.
\end{equation}
In this case, Eq. (31) reads as
\begin{equation}\label{034}
\bigtriangledown^{2}\ln\rho=\frac{4e}{\kappa^{2}}\rho^{2}.
\end{equation}
Let $\rho=\sqrt{f}$, then Eq. (\ref{034}) gives
\begin{equation}\label{035}
\bigtriangledown^{2}\ln{f}=\frac{8e}{\kappa^{2}}f,
\end{equation}
which is a nonlinear equation known as the Liouville equation, and
has the general exact solutions
\begin{equation}\label{036}
f=\frac{\kappa^{2}}{4e}\nabla^{2}\ln{(1+|g|^{2})},
\end{equation}
where $g=g(z)$ is a holomorphic function of $z=x^{1}+ix^{2}$. So
we obtain the general analytical solution of Eq. (\ref{034}):
\begin{equation}\label{037}
\rho=\sqrt{\frac{\kappa^{2}}{4e}\nabla^{2}\ln{(1+|g|^{2})}} \;,
\end{equation}
where  the corresponding charge density $\rho$ of the vortex must
obey the condition (\ref{033}), so the holomorphic function
$g=g(z)$ is not arbitrary.

\section{conclusion}
Our investigation is based on the connection between the self-dual
equation of Chern-Simons-Higgs model and the $U(1)$ gauge
potential decomposition theory and $\phi$-mapping theory. First,
we directly obtain one special form of $U(1)$ gauge potential
decomposition from the first of the self-duality equations.
Moreover, we obtain the inner topological structure of the
Chern-Simons vortex. The multicharged vortex has been found at the
Jacbian determinate $D(\phi/x)\neq0$. It is also showed that the
charge of the vortex is determined by Hopf indices and Brouwer
degrees. Second, we establish the rigorous self-duality equations
with topological term for the first time, in which the topological
term is the density of topological charge of vortex:
$4\pi\delta^{2}(\vec{\phi})D(\frac{\phi}{x})=2eJ^{0}$. In contrast
with the conventional self-duality equation (\ref{003}), one can
see that the self-duality equation with topological term is valid
when the field $\phi=0$; topological term vanishes and the
self-duality equation becomes Eq. (\ref{003}) when the field
$\phi\neq0$.
Additionally, the analytical vortex
solution of the equations in the special condition (\ref{033}) is
obtained, the charge density $\rho$ of the vortex must obey the
condition.

\section{Acknowledgements}
It is a pleasure to thank Dr. Lijie Zhang and Zhenhua Zhao for
interesting discussions. This work was supported by the National
Natural Science Foundation and the Doctor Education Fund of
Educational Department of the People's Republic of China.

\end{document}